# Low frequency noise due to magnetic inhomogeneities in submicron FeCoB/MgO/FeCoB magnetic tunnel junctions

D.Herranz[1], A. Gomez-Ibarlucea[1], M. Schäfers[2], A. Lara[1], G. Reiss[2], and F.G. Aliev[1*]

(1) Dpto. Física Materia Condensada, C03 Universidad Autónoma de Madrid, 28049 Madrid, Spain

(2) Dept. of Physics, University of Bielefeld, Bielefeld 33615, Germany

We report on room temperature low frequency noise due to magnetic inhomogeneities/domain walls (MI/DWs) in elliptic submicron FeCoB/MgO/FeCoB magnetic tunnel junctions with an area between 0.0245 and 0.0675μm$^2$. In the smaller area junctions we found an unexpected random telegraph noise (RTN1), deeply in the parallel state, possibly due to stray field induced MI/DWs in the hard layer. The second noise source (RTN2) is observed in the antiparallel state for the largest junctions. Strong asymmetry of RTN2 and of related resistance steps with current indicate spin torque acting on the MI/DWs in the soft layer at current densities below 5×10$^5$ A/cm$^2$.

(*) corresponding author: farkhad.aliev@uam.es

The discovery of large tunneling magnetoresistance (TMR) [1-6] has boosted interest in magnetic tunnel junctions (MTJs) which show reduced power consumption, high TMR and spin torque (ST) [7]. While MTJs with sizes of tens of microns are optimal for magnetic field detectors [8], junctions below 100nm are used for ST magnetic random access memories [9] or microwave oscillators [10]. Electron transport and *low frequency magnetic noise* in submicron MTJs of some hundreds of nanometers, where single magnetic inhomogeneities (MI) [11] and domain walls (DW) play an important role in magnetization reversal, remain poorly understood.

Previous studies of 1/f (magnetic, nonmagnetic and electronic noise) and random telegraph noise (RTN) focus on MTJs above-micron size with $Al_2O_3$ [12-14] and MgO [8,15-20] barriers. Recent advances in understanding of magnetic 1/f noise are summarized in [14,18,19]. Our letter presents both TMR and low frequency noise at room temperature in CoFeB/MgO/CoFeB MTJs with 0.8nm thick MgO barriers and areas from 0.117 $\mu m^2$ to 0.0245$\mu m^2$. MTJs of these sizes reveal two qualitatively different, robust and reproducible RTN types related with single MI/DWs. The resistance steps and related RTN found in antiparallel (AP) state in the largest MTJs are asymmetrically influenced by the current suggesting an influence of spin torque on MI/DWs at very small tunnel current densities.

The layer stack was deposited by magnetron sputtering in a Timaris PVD cluster tool from Singulus Technologies. Its structure is Ta5/Cu-N90/Ta5/Pt-Mn20/Co-Fe 2.2/Ru0.8/Co-Fe-B2/Mg0.8 +1200s oxidation+Mg 0.3/Co-Fe-B 2/Ta 10/Cu-N 30/Ru-7 (thicknesses in nm). The stack was annealed for 90 minutes at 360°C and cooled in a field of 1 Tesla to establish the exchange bias. Using electron beam lithography and ion beam milling the stack was patterned into elliptic tunnel junctions with different sizes from 600nm×250nm to 260nm×120nm. The zero bias TMR was between 45% and 160% and R×A (Resistance-Area products) between 3Ω×μm$^2$ and 19Ω×μm$^2$. Out of 13 MTJs with TMR at room temperature exceeding 45% (Fig.1a) we present low frequency (1Hz-10 kHz) noise measurements for 7 MTJs which reversibly stood biases between 100mV and 400mV. We measured the power spectrum $S_V(f)$ [15] and TMR along the easy (elliptic) axis. To compare the noise level in different junctions, we use Hooge factor (α) from: $S_v(f)=\alpha \times V^2/(A \times f)$, where A is the area, f is frequency and V the bias. Strong deviations from this dependence are usually caused by RTN [21]. Following arguments rule out RTN explanation in terms of pinholes or "hot spots": (i) large TMR, (ii) robustness of the MTJs to bias and multiple field scans, and (iii) low values of the Hooge factor in the saturated P state ($3.1\times10^{-11}$ μm$^2$), close to expected from the empirical summary [22].

Multiple slow field sweeps between the AP and parallel (P) states minimize metastable states and lead, for most of the MTJs, to two types of noise behavior depending on the junction area. Most of the smallest junctions (0.0245μm$^2$-0.0503μm$^2$) reveal enhancement of the noise in the P state (Fig.1b) which has a RTN origin (further RTN1). Very strong increase of the noise power (Fig.1b) is accompanied by a change in the resistance of less than 0.2%. Estimations show that changes of RTN1 with bias polarity (Fig.1b) could be partially due to self-field.

To estimate the fluctuating magnetic moment (Δm) we have measured field and bias dependence of the ratio between inverse attempt transition rates [12,13]: $\tau_1/\tau_2 \sim \exp(-2\Delta mH/k_BT)$, $\tau_1$ and $\tau_2$ being the average time spent in each of two activated states. This relation considers the magnetization directions of fluctuating states being P-AP to the external field. Figure 1c shows typical histograms of RTN1 for magnetic field above and below the maximum of noise in P state. A linear fit of $\ln(\tau_1/\tau_2)$ vs. H (Fig.1d) provides an estimation of the fluctuating moment of $1.5\times10^5\mu_B$ and within 10% being independent of the bias polarity up to 400 mV. We estimate that DW/MI occupy about 10% of the soft electrode area (with CoFeB moment per atom of $1\mu_B$ [23]). To account for maximum 0.2% variation of the resistance in the P state near RTN1, the related DW/MI should be located in the hard layer *outside* the MTJ stack.

The largest submicron MTJs (A=$0.0565\mu m^2$, $0.0675\mu m^2$) do not show noise anomalies in the P state but reveal (Fig.3a) a strong noise enhancement in the AP state at least 100 Oe above the AP-P transition, also originated from RTN (further RTN2). The investigation of the RTN2 time series as a function of magnetic field provides fluctuating moment of $4\times10^5\mu_B$. New feature of the RTN2 is that it exists only with positive bias corresponding to the injection of electrons from the hard to the soft electrode (Fig.2a,b). The MI/DWs which originate RTN2 are probably located in the free electrode. Indeed, its estimated fluctuating area is about 7% **,** in rough agreement with 2% reduction of TMR in the AP state (Fig.2a).

The absence of RTN2 for negative biases indicates possible influence of spin torque on DW/MI. If the positive bias favors the P alignment, it will destabilize the AP alignment, while the negative bias direction would favor an AP alignment of both electrodes and suppress RTN2. Unlike RTN1, fluctuations similar to RTN2 were seen in GMR nanopillars [24-26], but at current densities above $10^7$ A/cm$^2$.

To estimate self-fields $H_{self\text{-field}}$ we assume circular nanopillars with uniform current density J. The surface integral of the J crossing surface $\sigma$ equals the line integral of the self-field $H_{self\text{-field}}$ along $\sigma$ contour, $\partial\sigma$: $\oint_{\partial\sigma}\vec{H}d\vec{l} = \iint_{\sigma}\vec{J}d\vec{S}$. For a circular path $\partial\sigma$ of radius $r$ this provides $H_{self-field} = |\vec{J}|\times r/2$ of few hundreds of Oe for GMR nanopillars [25]. To verify this for RTN2, Fig.3a compares estimated $H_{self\text{-field}}$ with

dependence of the soft layer coercive field ($H_c$) and the fields where resistance steps and RTN2 in the AP states are observed ($H_{RTN2}$) (both referenced to $H_c$ at zero bias) as a function of J. This analysis rules out significant influence of self-field on RTN2. We further checked the effects of self-fields on RTN2 kinetics by attempting to compensate them with changes in magnetic field. Figures 3(b-d) show that to compensate changes related to current with estimated variation of self-field below 2 Oe one should vary external field in about 30 Oe. This indicates that the effects of self-fields are not sufficient to explain the asymmetry in RTN2 kinetics.

A simple model qualitatively explains the possible origin of the RTN(1,2) (Fig.4). While the soft electrodes in the smallest MTJs remain in a single domain state, close to magnetization inversion, the largest electrodes show, for the same fields, (independently of anisotropy [27] as confirmed by simulations) DW/MI formation. A 7% reduction of the magnetization of the soft layer (Fig.4b), which could provide RTN2, is observed. Formation of the small ($10^5 \mu_B$) DW/MI2 in the AP state could affect the current distribution (which is mainly of $\Delta_5$ symmetry for the ideal AP alignment [1,2]) creating "pseudo-pinhole" for electrons with $\Delta_1$ symmetry (Fig.4a). This spin current excess, alongside large perpendicular ST [7], could explain the influence of spin current on RTN2 already at current densities, below $10^6$ A/cm$^2$, which are at least a factor of 10 smaller than for GMR nanopillars [24-26].

The RTN1 is most probably due to DW/MI1 located in the biased layer outside and close to the edge of the MTJ pillar (Fig.4a). The origin of DW/MI1 could be $360^0$ DWs [28] pinned by the stray field of the soft layer (Fig.4a). Figure 4b summarizes the characteristic magnetic fields where reproducible maxima of RTN(1,2) were observed. The absence of RTN1 in largest MTJs with reduced influence of the edge stray field (Fig.4b) contradicts explanation the RTN1 due to defects in the MgO influenced by magnetostriction.

Authors thank J.P.Cascales for technical assistance. The work was supported by Spanish MICINN (MAT2009-10139, CSD2007-00010) and CAM (P2009/MAT-1726).

Figure captions

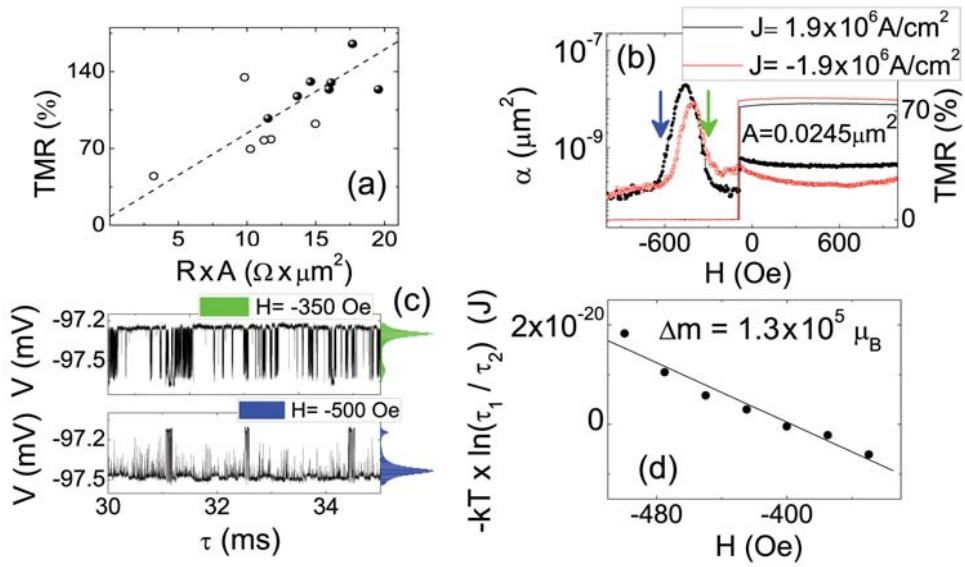

Figure 1 a) Normalized TMR vs. R×A. Dashed line is guide for the eyes. Closed dots indicate MTJs for which noise measurements have been done.

(b) TMR (line) and Hooge factor (points) in MTJ of 0.0245μm$^2$ for opposite current densities ±1.9×10$^6$ A/cm$^2$. Arrows correspond to the fields for which time traces are presented in part (c).

(c) Time series and corresponding histograms measured in two fields above and below of maximum in noise indicated by arrows in part (b).

(d) Logarithm of the relation between inverse attempt transition rates as a function of magnetic field. Solid line is mean square fit.

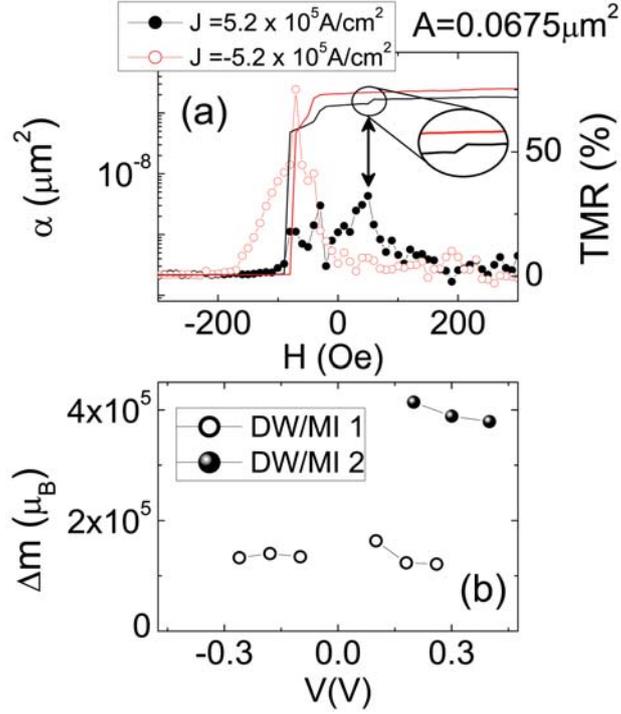

Figure 2 (a) TMR (line) and Hooge factor (points) in MTJ with area of 0.0675μm$^2$ measured with opposite current densities of ±5.2×10$^5$ A/cm$^2$.

(b) Comparative bias dependences of estimated fluctuating moments obtained from RTN1 and RTN2.

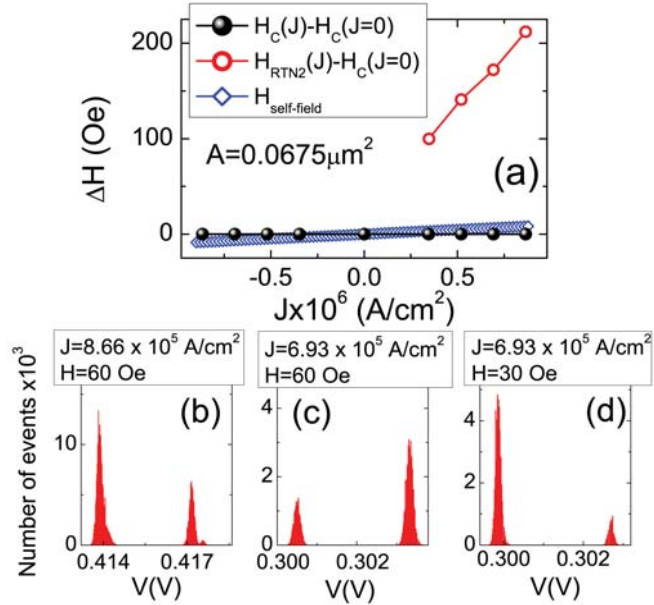

Figure 3 (a) Estimated maximum self-field (H$_{self-field}$) in comparison with dependence of coercive field (H$_c$) of the soft layer, and characteristic field (H$_{RTN2}$) where resistance steps and RTN2 in the AP state are observed (referenced to $H_c$ at zero bias) as a function of applied current density. Parts (b-d) show that variation of RTN2 under change in bias current (with estimated change in self-field below 2 Oe) is roughly compensated by external magnetic field of 30 Oe

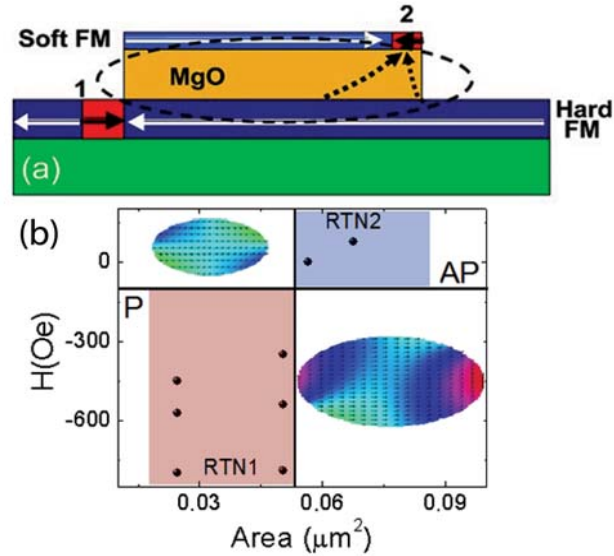

Figure 4 (a) Sketch of MTJs. DW/MIs inside hard and soft electrodes indicated as (1) and (2) respectively. Dotted lines inside MgO spin current with $\Delta_1$ symmetry into DW/MI2 in the AP state. Dashed line sketches stray field.

(b) Characteristic magnetic fields of RTN(1,2) as a function of area. The inset shows simulated with OOMMF [29] the small (0.0245 μm$^2$ – left) and largest (0.0675 μm$^2$- right) soft electrodes corresponding to 7% reduction of magnetization in the larger dot due to appearance of DW/MI2. Parameters used are: saturation magnetization of $1150\times10^3$ A/m, exchange stiffness of $2\times10^{-11}$ J/m and magnetization damping 0.01. Qualitatively similar results were obtained in the presence of uniaxial anisotropy $K_u$=1990J/m$^3$ [27].